\documentclass[prb,twocolumn,aps,showpacs,superscriptaddress,epsfig]{revtex4-2}
\usepackage{dcolumn}
\usepackage{bm}
\usepackage[mathlines]{lineno}
\usepackage{amssymb}
\usepackage[usenames,dvipsnames]{xcolor}
\usepackage{amsmath}
\usepackage{amstext}
\usepackage{latexsym}
\usepackage{amsfonts}
\usepackage{graphicx,epsfig,psfrag,subfigure,amsopn,epstopdf}
\modulolinenumbers[5]
\newcommand\sg{\subfigure}
\newcommand\ig{\includegraphics}
\DeclareMathOperator{\Tr}{Tr}
\begin{document}
\title{Insights from the analytical solution of a periodically driven transverse field Ising chain}
\author{Pritam Das}
\author{Anirban Dutta}
\email{anirband@bhu.ac.in}
\affiliation{Department of Physics, Institute of Science, Banaras Hindu University, Varanasi- 221005, India}
\date{\today}
\begin{abstract}
We derive an exact analytical expression, at stroboscopic intervals, for the time-dependent wave function of a class of integrable quantum many-body systems, driven by the periodic delta-kick protocol. To investigate long-time dynamics, we use the wave-function to obtain an exact analytical expression for the expectation values of the defect density, magnetization, residual energy, fidelity, and the correlation function after $n$th drive cycle. Periodically driven integrable closed quantum systems absorb energy, and the long-time universal dynamics are described by the periodic generalized Gibbs ensemble(GGE). We demonstrate that the expectation values of all observables are divided into two parts: one highly oscillatory term that depends on the drive cycle $n$, and the rest of the terms are independent of it. Typically, the $n$-independent part constitutes the saturation at large $n$ and periodic GGE. The contribution from the highly oscillatory term vanishes in large $n$. We also generalize our formalism to include square pulse and sinusoidal driving protocols.

\end{abstract}
\maketitle
\section{Introduction}
In recent years, significant experimental\cite{exp1,exp2,exp3,exp4,exp5} advancements in the closed quantum systems have also deepened our theoretical \cite{Dziarmaga,ks_rmp,polkov_rev,rigol_rev,bkc_book,iop_rev,saito_rev,CZ12} understanding of the driven quantum systems, which is critical for progress in foundational quantum mechanics, quantum computing, and quantum information theory. A variety of experimental setups have been developed, including ultracold atoms, ion traps, semiconductor qubits, and superconducting qubits, to test the theoretical predictions rigorously. Initial research in the field of driven quantum systems focused on the sudden quench and ramp protocols, where the density of excitation at the final state of the drive follows the universal Kibble-Zurek scaling. While such quench protocols are still being studied, there is a growing interest in driving the closed quantum systems out of equilibrium using a periodic external perturbation\cite{kr_floq_rev,fazio_rev,eckardt_rev,gritsev_rev}. The ability to modify and control material properties using optical pulses has gathered significant interest from the scientific community due to its potential application in technology\cite{Experiment2,Experiment1,exp_femto,exp_ultras,expt_FH,nori_he}. The Floquet theory makes it easier to describe such systems at discrete time intervals by using a time-independent Floquet Hamiltonian derived from the time evolution operator over one time period. This approach has resulted in the emergence of Floquet engineering as a novel research area.  Advances in the Floquet engineering have enabled the preparation of systems in different equilibrium phases and uncovered new non-equilibrium phases and also the transitions between them. Recently, Floquet theory reveals a number of distinct phenomena in these systems that do not occur in equilibrium or other driven systems. Dynamical localization\cite{dl1,dl2,dl3} and freezing\cite{df1,df2,df3,df4,df5}, dynamical transitions\cite{nori_rev,dt1}, Floquet topological phases\cite{topo1,topo3,topo2,topo5,topo4,A.D5}, Floquet realization of quantum scars\cite{Bhaskar,bm1,bm2,bm3,ks1}, time-crystalline states of matter, and the optically controlled materials, are just a few examples. These phenomena make the periodically driven systems a dynamic field of research, yielding new insights into how time-periodic drives can fundamentally alter the system's behavior and properties.\\
In this study, we investigate the nonequilibrium dynamics of a periodically driven quantum integrable system using a semianalytical method. While most analytical approaches for the periodically driven systems rely on the Floquet theory, which assumes an infinite duration for the periodic driving, (i.e., $n\to \infty$), this is not practical in experimental settings. Nevertheless, it is expected that for sufficiently large $n$, experimental results should match with those predicted by the Floquet theory. Recent advances in the ultrafast laser science have provided a unique opportunity to investigate the Floquet engineering with extremely high time resolution, allowing one to study the periodically driven states on shorter timescales\cite{exp_femto,exp_ultras}. For finite $n$, most of the calculations are typically performed numerically. Our goal is to bridge this gap by deriving an analytical expression for the time-dependent wavefunction of a periodically driven system after finite $n$ drive cycles. Our results are exact and valid for any finite $n$ and drive frequency $\omega$, enables experimentalists to verify their findings without relying on Floquet theory. Additionally, we provide an analytical expression for the expectation values of the several widely studied observables found in the literature, after $n$ periods. To our knowledge, this kind of extensive analytical exact expression for periodically driven quantum many-body systems has not been reported before.\\
This formulation applies to any experimentally realizable system that can be simplified to a quantum two-level system. It can also be extended to a class of exactly solvable models that can be expressed in the following form:
\begin{eqnarray}
H&=&\sum_{k}\psi_{k}^{\dagger}\left[(g(t)-a_{k})\sigma^3+b_k\sigma^1\right]\psi_{k},
\end{eqnarray}
where, example Hamiltonian includes models such as the transverse field Ising (TFI) chain, the quantum XY chain, the Kitaev chain in one dimension(1D), and the Kitaev model in 2D. Here, $a_k$ and $b_k$ are functions of momentum only, and $g(t)$ is the time-dependent parameter varied according to a time-periodic protocol. $\sigma^1$ and $\sigma^3$ are Pauli matrices, and $\psi_{k}^{\dagger} = (c_{k}^{\dagger}, c_{-k})$, where $c_{k}$ is the fermionic annihilation operator in momentum space, given by
\begin{eqnarray}
c_{k}&=&\frac{1}{\sqrt{L}} \sum_{j=1}^{L} e^{ikj} c_j,
\end{eqnarray}
$c_j$ is fermion annihilation operator at site $j$ and $L$ is the total number of sites. In this paper we will focus solely on the 1D TFI chain for clarity.\\
Although this formulation can be applied to any periodic driving protocol, provided the evolution operator for one time period can be exactly known analytically, we restrict our calculations to the periodically kicked driving protocol for simplicity of the calculation. In the appendix, we have included the calculations with the square-pulse driving protocol. For a sinusoidal driving protocol, it is possible to either find the analytical expression for the evolution operator over one time period using the adiabatic-impulse-adiabatic approximation or compute it numerically by solving the time-dependent Schrödinger equation. In Sec.~\ref{sec:sinu}, we numerically find the evolution operator for one time period and extend our formulation to sinusoidal driving protocols. Additionally, this approach can also be adapted to some non-Hermitian systems, which have recently attracted significant interest\cite{nhrev}.\\
The main results of our study are as follows. We derive an exact analytical expression for the time-dependent wavefunction after $n$th drive cycles. Using this expression, we obtain an analytical formula for the expectation value of the defect density produced after $n$ drive cycles. Additionally, we calculate the analytical expressions for the residual energy, magnetization, and fidelity. Furthermore, we provide the exact expression for all the nontrivial two-point correlators. Using these results, we numerically compute the entanglement entropy, which is consistent with the results available in the literature. Our expressions have streamlined the numerical calculations and reduced the time needed for determining the entanglement entropy and other observables for the driven system discussed here.\\
The paper is organized as follows: In Sec. \ref{sec_ham}, we provide a brief introduction to the system of interest, the transverse field Ising (TFI) chain Hamiltonian, and review some of its key properties. We also introduce the periodic driving protocol and derive the analytical expression for the stroboscopic time-dependent state after any arbitrary $n$ drive cycles. Using this expression, we derive the formula for the residual energy in Sec. \ref{sec_dyn_loc} and discuss its properties. In Sec. \ref{sec_dyn_fre}, we derive the expression for magnetization and explore the concept of dynamical freezing. Next in Sec. \ref{sec_fide} we present the expression for fidelity and discuss properties. We discuss all two-point correlators in Sec. \ref{sec:corr}. Before concluding, in Sec.~\ref{sec:sinu}, we investigate the sinusoidal driving by numerically computing the time evolution operator for a single time period.  We then apply our formalism to calculate the expectation values of similar observables after any finite $n$ drive cycle. Finally, we discuss our results and conclude in Sec. \ref{sec_diss}. Details of the calculations for determining the $n$th power of the evolution matrix are provided in the Appendix \ref{sec_app}. In Appendix \ref{sec_app_sq}, we apply our formalism to the periodic square-pulse driving protocol.
\section{Model Hamiltonian}
\label{sec_ham}
In this section, we will review the exactly solvable model required for our discussion on the periodic driving. To simplify calculations, we have chosen the one-dimensional free-fermionic TFI chain for this paper. We have discussed some of the important properties of the model followed by the details of our driving protocol. Next we provide an analytic expression for the driven wavefunction, followed by an analytic expression for the defect density. 
\subsection{Transverse-field Ising chain Hamiltonian}
To begin, let us briefly discuss the model: a spinless, free fermionic one-dimensional TFI chain with $L$ sites. The time-independent Hamiltonian of this model is given by:
\begin{eqnarray}
H = -J \sum_{i=1}^{L} \left( \sigma_i^x \sigma_{i+1}^x + g_0 \sigma_i^z \right),
\end{eqnarray}
where, $\sigma_i^\alpha$ denotes the Pauli spin matrices at the site $i$, $J$ is the strength of the nearest-neighbor Ising interaction, and $g_0$ is the strength of the transverse field. The Ising Hamiltonian can be transformed into a quadratic free fermionic Hamiltonian using the nonlocal Jordan-Wigner transformation:
\begin{eqnarray}
\sigma_i^z &=&1 - 2 c_i^\dagger c_i, \\
\sigma_i^x &=&- (c_i + c_i^\dagger) \prod_{j<i} (1 - 2 c_j^\dagger c_j),
\end{eqnarray}
where, $c_i$ is the fermionic annihilation operator at the site $i$. Utilizing translational invariance, the quadratic fermionic Hamiltonian in the momentum space can be expressed as $H = \sum_{k} \psi_k^\dagger H_k \psi_k$, with:
\begin{eqnarray}
H_k = (g_0 - \cos k) \sigma^3 + (\sin k) \sigma^1,
\end{eqnarray}
and $k = \pm \frac{1}{2} \frac{2\pi}{L}, \ldots, \pm \frac{L-1}{2} \frac{2\pi}{L}$. For simplicity, we set $J = 1$ for all calculations, and all comparisons are made in units of $J$. The total Hilbert space of the Hamiltonian is a product of two-dimensional subspaces spanned by $|0_k, 0_{-k}\rangle$ and $|1_k, 1_{-k}\rangle$ for each $k$ mode. The Hamiltonian in momentum space can be diagonalized using the Bogoliubov transformation:
\begin{eqnarray}
c_k = u_k \gamma_k + v_{-k}^* \gamma_k^\dagger.
\end{eqnarray}
The final Hamiltonian in the Bogoliubov basis can be written as:
\begin{eqnarray}
H = \sum_{k > 0} \epsilon_k \left( \gamma_k^\dagger \gamma_k - \frac{1}{2} \right),
\end{eqnarray}
where, the eigenvalues are $\pm \epsilon_k$, with:
\begin{equation}
\epsilon_k = \sqrt{(g_0 - \cos k)^2 + \sin^2 k},
\label{eq:dis}
\end{equation}
and the eigenvectors are as follows:
\begin{equation}
\left( \begin{array}{c} u_k \\ v_k \end{array} \right) = \left( \begin{array}{c} \cos \frac{\phi_k}{2}\\ \sin \frac{\phi_k}{2} \end{array} \right),
\end{equation}
where, $\tan \phi_k = \frac{\sin k}{g_0 - \cos k}$. The system becomes gapless at $k = 0$, $g_0=1$ or $k = \pi$, $g_0=- 1$, indicates the presence of a quantum critical point\cite{Subir} that separates the ferromagnetic and the paramagnetic phase. The Hamiltonian can be made time-dependent by allowing $g_0$ to vary with time, i.e., $g_0 = g(t)$. For a time-dependent parameter $g(t)$, the Hamiltonian in the momentum space is given by:
\begin{equation}
H_k(t) = (g(t) - \cos k) \sigma^3 + (\sin k) \sigma^1.
\end{equation}
The nonequilibrium dynamics of the model can be studied by solving the time-dependent Schrödinger equation
\begin{equation}
i\partial_t |\psi_k\rangle = H_k(t) |\psi_k\rangle
\end{equation}
for each mode $k$. We are particularly interested in a special category of driven systems where the parameter $g(t)$ varies periodically over time, that is, $g(t)=g(t + T)$, where $T$ is the period of the driving protocol. We further simplify the driving protocol by assuming the transverse field is periodically modulated by delta pulse kicks with frequency $\omega$, given by
\begin{equation}
g(t)=g_0+g_1\sum_{s=-\infty}^{\infty}\delta(t-sT),
\label{eq:kick}
\end{equation}
where $T = \frac{2\pi}{\omega}$ and $s$ is an integer. The periodic driving makes the Hamiltonian time-dependent, such that $H(t + T)=H(t)$. The stroboscopic time evolution operator over one period for such a time-periodic Hamiltonian, can be written as
\begin{equation}
U_k(T) = \mathbb{T} \exp \left( -i \int_0^T H(t) \, dt \right) = \exp \left( -i H_k^F T \right),
\end{equation}
where $H_k^F$ is the time-independent Floquet Hamiltonian and $\mathbb{T}$ is the time ordered product. Calculating the Floquet Hamiltonian is generally quite difficult because of the complex time-ordering involved. As a result, the majority of research on the periodically driven systems has concentrated on obtaining the Floquet Hamiltonian through various approximation methods. It is important to note that our calculations do not rely on the Floquet Hamiltonian and are also applicable in cases where the analytic expression for the Floquet Hamiltonian is unknown. For the delta-kicked driving protocol, the symmetrized time-evolution operator is given by\cite{Manisha,Amit_Dutta}
\begin{equation}
U_k(T)=e^{-i\frac{g_1 }{2}\sigma^3} e^{-i T \left[ (g_0 - \cos k) \sigma^3 + (\sin k) \sigma^1 \right]} e^{-i \frac{g_1}{2} \sigma^3},
\end{equation}
which simplifies to
\begin{eqnarray}
U_k(T)=\begin{pmatrix}
\alpha_k & -\beta_k^* \\
\beta_k & \alpha_k^*
\end{pmatrix},
\label{eq:uk}
\end{eqnarray}
where,
\begin{eqnarray}
\alpha_k&=&e^{-i g_1} \left( \cos(\Phi_k) - i n_{kz} \sin(\Phi_k) \right),\\
\beta_k&=&-i n_{kx} \sin(\Phi_k),\\
\Phi_k&=&T \epsilon_k,\\
n_{kz}&=&\frac{g_0 - \cos k}{\epsilon_k},\\
n_{kx}&=&\frac{\sin k}{\epsilon_k}.
\label{eq:delta_para}
\end{eqnarray}
Eq. (\ref{eq:uk}) is the starting point of extending our calculation to other driving protocols.
\subsection{Time-dependent state after $n$th period}
In this subsection, we derive the time-dependent state after $n$th period, where $n$ is a positive integer. Using the time-evolution operator, one can determine the wave function after $n$ periods as,
\begin{eqnarray}
|\psi_k(t = nT)\rangle = U_k^n(T) |\psi_k(t = 0)\rangle.
\end{eqnarray}
With an analytical expression for the evolution operator over one time period, one can easily compute the $n$th power of the $2\times2$ matrix after a few lines of algebra (see details in the Appendix A). Using the properties of SU(2) matrix, the $n$th power of the time evolution operator is given by\cite{Griffiths}
\begin{eqnarray}
U_k^n(T) &=& U_k(T) \cdot \mathcal{F}_{n-1}(\xi_k) - \mathbb{I}_2 \cdot \mathcal{F}_{n-2}(\xi_k),\\
\xi_k &=& \frac{1}{2} \text{Tr}(U_k(T)) = \mathfrak{Re}[\alpha_k],
\end{eqnarray}
where $\mathcal{F}_{n}(x)$ denotes the Chebyshev polynomials of the second kind of degree $n$ in $x$, and $\mathbb{I}_2$ is the $2\times2$ identity matrix. The time-dependent state after $n$th drive periods can be expressed in terms of the initial state and the other Hamiltonian parameters as
\begin{widetext}
\begin{eqnarray}
\left( \begin{array}{c}
u_k(n) \\
v_k(n)
\end{array} \right)&=&
\left(\begin{array}{cc}
\alpha_k \mathcal{F}_{n-1}(\xi_k) - \mathcal{F}_{n-2}(\xi_k) & -\beta_k^* \mathcal{F}_{n-1}(\xi_k) \\
\beta_k \mathcal{F}_{n-1}(\xi_k) & \alpha_k^* \mathcal{F}_{n-1}(\xi_k) - \mathcal{F}_{n-2}(\xi_k)
\end{array} \right)
\left( \begin{array}{c}
u_k(0) \\
v_k(0)
\end{array} \right).
\label{eq:nth}
\end{eqnarray}
\end{widetext}
The above Eq.(\ref{eq:nth}) constitutes closed form expression for the time-dependent state at stroboscopic interval. For simplicity, let us consider the initial state is a product of, $|\psi_{k}(t=0)\rangle=\left(0,1\right)^{\text{T}}$. While an analytic solution can be found for any initial state, this particular initial state simplifies the calculations further. In this limit, $u_k(n)$ and $v_k(n)$ simplify to,
\begin{eqnarray}
u_k(n) &=& -\beta_k^* \mathcal{F}_{n-1}(\xi_k) = -i n_{kx} \sin \Phi_k \frac{\sin n \theta_k}{\sin \theta_k},\label{eq:nth1}\\
v_k(n) &=& \alpha_k^* \mathcal{F}_{n-1}(\xi_k) - \mathcal{F}_{n-2}(\xi_k)\nonumber\\
&=& \cos n \theta_k + i \mathfrak{Im}[\alpha_k^*] \frac{\sin n \theta_k}{\sin \theta_k}.\label{eq:nth2}
\end{eqnarray}
Here, we have used the property of Chebyshev polynomials of the second kind: $\mathcal{F}_{n}(x)=\frac{\sin(n+1)\theta}{\sin\theta}$ with $x=\cos\theta$. Note that $\theta_k$, appearing in the equation
\begin{equation}
\theta_k = \cos^{-1} \xi_k = \cos^{-1}(\cos g_1 \cos \Phi_k - n_{kz} \sin g_1 \sin \Phi_k), \label{eq:thetk}
\end{equation}
represents the eigenvalues $\exp(i \theta_k)$ of the evolution operator $U_k(T)$ given in Eq.(\ref{eq:uk}). Equations (\ref{eq:nth1}) and (\ref{eq:nth2}) constitute the significant analytical results that have been verified with exact numerics. Almost similar results for two-level system are also obtained in\cite{nori_rev,Nori1}. In the rest of this paper, we will use these expressions to provide analytical expressions for the expectation values of several commonly studied observables found in the literature. Finally, before conclusion, please note that obtaining exact and analytical solutions for a periodically driven integrable many-body system in arbitrary parameter regimes remains challenging and is often only feasible through exact numerical methods. The Jordon-Wigner transformation in the TFI model decouples the entire Hilbert space into independent $k$ modes, which simplifies our task.
\subsection{Defect density}
\begin{figure}[t]
\begin{center}
\sg[~]
{\includegraphics[width=0.494\columnwidth]{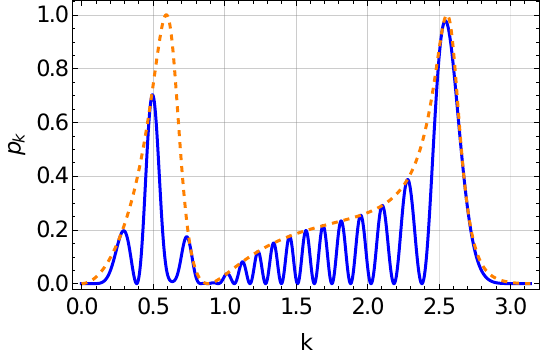}\label{fig:1a}}
\sg[~]
{\includegraphics[width=0.494\columnwidth]{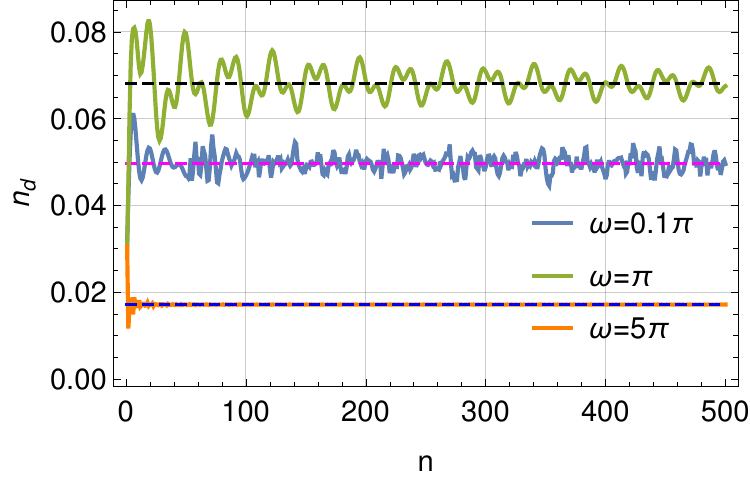}\label{fig:1b}}
\end{center}
\caption{(a) The behavior of $|u_k(n)|^2$ as a function of $k$ for $n=15$, plotted as a solid blue line. Higher $n$ corresponds to more oscillations. The slowly varying function $\Gamma_k^2/2$ is shown as a dashed orange line. (b) The  defect density generated after $n$ drive cycles is plotted against $n$ for different frequencies. The dashed lines represent the saturation value for large $n$, computed using Eq.~(\ref{eq:ndsa}). Parameters used: $L=2048$, $g_0=2$, $g_1=0.5$.}
\label{fig:1}
\end{figure}

In this subsection, we first calculate the excitation probability, which is essentially the overlap of the time-dependent state with the excited state of the two-level system at $t = 0$ for each value of $k$. The defect density for the TFI model is the sum of the excitation probabilities over each $k$ mode, divided by the total number of sites. For quench and ramp protocol, defect density at final state shows universal Kibble-Zurek scaling\cite{Kibble76,Zurek96,A.D1,A.D3}. Using the exact wave function, the excitation probability after $n$th drive cycles\cite{floq_kz,A.D2} is defined as
\begin{eqnarray}
p_k(n)=|\langle (1,0)|\psi_k(n)\rangle|^2=|u_k(n)|^2.
\end{eqnarray}
This is given by
\begin{eqnarray}
|u_k(n)|^2 = \frac{\Gamma_k^2}{2} \left(1 - \cos 2n \theta_k \right);~\quad \Gamma_k = \frac{n_{kx} \sin \Phi_k}{\sin \theta_k},
\label{eq:ukn}
\end{eqnarray}
where $\Gamma_k$ depends on the $n_{kx}, \Phi_k$ and $\theta_k$ as defined in the previous section. The above expression is a product of two terms: one highly oscillatory term $\cos 2n \theta_k$ coming from the Chebyshev polynomials of order $n-1$, and the other term, $\Gamma_k$, which varies slowly with $k$. We have plotted this expression in Fig. (\ref{fig:1a}) as a function of $k$ for $n=15$, with the slowly varying term $\Gamma_k$ shown in orange, serves as an envelope. The number of drive cycles $n$ influences the oscillatory behavior, with a higher $n$ leading to more oscillations within the envelope. The wave function after the $n$th drive cycles is related to the Chebyshev polynomials of degrees $n-1$ and $n-2$. Since the Chebyshev polynomial of order $n$ has $n$ roots, increasing $n$ results in more oscillations in the wave function. Similarly, one can find an expression for $|v_k(n)|^2$, as the sum of $|u_k(n)|^2$ and $|v_k(n)|^2$ must add up to unity to maintain normalization.
Using Eq.~(\ref{eq:ukn}), we can also calculate the defect density $n_d(n)$ generated after $n$ drive cycles, given by
\begin{equation}
n_d(n) = \frac{1}{L} \sum_{k > 0} p_k(n) = \frac{1}{L} \sum_{k > 0} \frac{\Gamma_k^2}{2} \left(1 - \cos 2n \theta_k \right).
\label{eq:def_def}
\end{equation}
This sum cannot be evaluated analytically, so we resort to numerical evaluation. Figure~\ref{fig:1b} shows the defect density $n_d$ as a function of $n$ for different frequencies, with the defect density saturating at large $n$. The saturation value can be calculated using the slowly varying function $\Gamma_k^2/2$ and is given by
\begin{equation}
n_d^{sat} = \frac{1}{L} \sum_{k > 0} \frac{\Gamma_k^2}{2} = \frac{1}{L} \sum_{k > 0} \left(\frac{n_{kx} \sin \Phi_k}{\sin \theta_k}\right)^2.
\label{eq:ndsa}
\end{equation}
Note that the only terms containing the drive frequency $\omega$ in the above expression are $\Phi_k=\frac{2 \pi \epsilon_k}{\omega}$ and $\theta_k$ that involves $\Phi_k$. At large $\omega$, $\sin^2 \Phi_k$ approaches zero, $\theta_k\sim g_1$, implying that $n_d^{sat}$ becomes nearly zero. Physically, this means that at high frequencies, the system fails to respond to the external drive and remains close to the initial state, resulting in zero defect. Additionally, at $g_1 = 0$ and $\omega = 2\pi$, i.e., $T=1$, we have $\theta_k=\Phi_k=\epsilon_k$, which simplifies the saturation value of the defect density to
\begin{equation}
n_d^{sat} = \frac{1}{L} \sum_{k > 0} \frac{1}{1 + \left(\frac{g_0 - \cos k}{\sin k}\right)^2} = \frac{1}{4 g_0^2}.
\label{sukn}
\end{equation}
\section{Residual energy}
\label{sec_dyn_loc}
\begin{figure}[htb]
\begin{center}
\sg[~]
{\includegraphics[width=0.494\columnwidth]{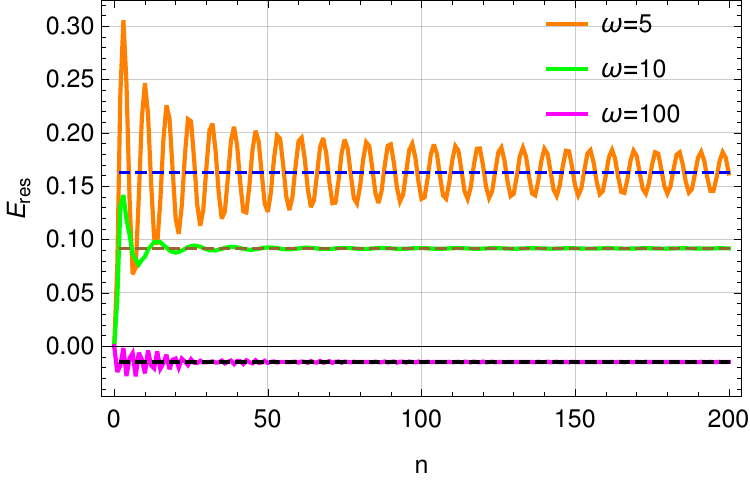}\label{fig:2a}}
\sg[~]
{\includegraphics[width=0.494\columnwidth]{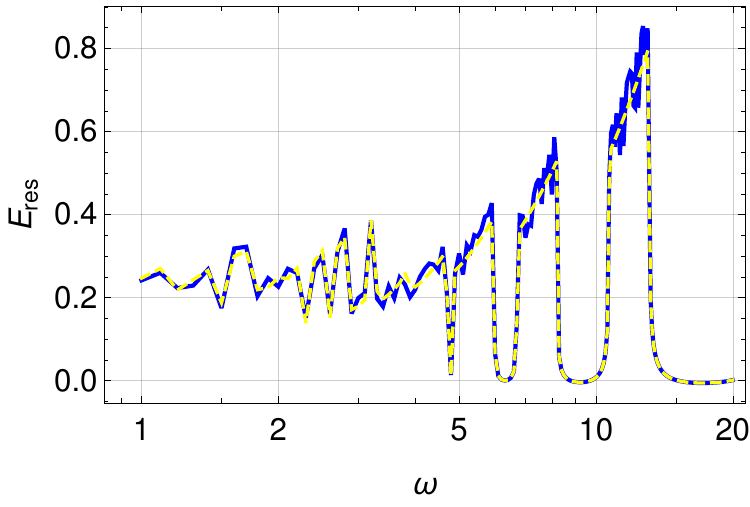}\label{fig:2b}}
\end{center}
\caption{(a) The characteristics of residual energy as a function of $n$. The large $n$ saturation value is indicated by the dashed line. The frequencies used are given in the inset. The parameters used in this figure are $g_0=2$ and $g_1=1$. (b) The residual energy at $n=500$ as a function of drive frequency. The dashed line corresponds to the saturation value given by Eq.~(\ref{eq:res_sat}). The parameters used in this figure are $g_0=10$ and $g_1=1$.}
\label{fig:2}
\end{figure}

In the previous section, we derived the expression for the defect generated due to energy absorption from the external drive. In this section, we focus on the residual energy after $n$ drive cycles, which determines the amount of energy absorbed by the system from an external drive. The residual energy is defined as the additional energy per site in the time-dependent state as compared to the ground state. Understanding the behavior of the residual energy is also crucial as it displays universal scaling of quantum systems near the critical point. Additionally, for periodically driven integrable systems, it has been shown that the zeros of the rate function in the work distribution statistics are related to the residual energy, enabling more accurate experimental measurements\cite{A.D2}.\\
We define the residual energy as
\begin{eqnarray}
E_{res}&=&\frac{1}{L}\sum_{k}\left(\epsilon_k(n)-\epsilon_k(0)\right)\nonumber\\
&=&\frac{4}{L}\sum_{k>0} \Big(|u_k(n)|^2 (g_0-\cos k)+\mathfrak{Re}(u_k^*(n)v_k(n))\sin k\Big)\nonumber\\,
\label{eq:res_def}
\end{eqnarray}
where $\epsilon_k(n)=\langle \psi_k(nT) | H_k^0 | \psi_k(nT) \rangle$ and $H_k^0$ is the initial Hamiltonian. Using the expression of wavefunction from the Eq.~(\ref{eq:nth}), analytical expression for the residual energy and its large $n$ saturation value are given by
\begin{widetext}
\begin{eqnarray}
E_{res}(n) &=& \frac{1}{L} \sum_{k > 0} \left[ \Gamma_k^2 (g_0 - \cos k) + \Gamma_k^2 \left( \sin g_1 \cos \Phi_k + n_{kz} \cos g_1 \sin \Phi_k - g_0 + \cos k \right) \cos 2n \theta_k \right] \nonumber \\
&& - \frac{\Gamma_k \sin k}{\sin \theta_k} \left( \sin g_1 \cos \Phi_k + n_{kz} \cos g_1 \sin \Phi_k \right),
\label{eq:resi} \\
E_{res}^{sat} &=& \frac{1}{L} \sum_{k > 0} \left[ \Gamma_k^2 (g_0 - \cos k) - \frac{\Gamma_k \sin k}{\sin \theta_k} \left( \sin g_1 \cos \Phi_k + n_{kz} \cos g_1 \sin \Phi_k \right) \right].
\label{eq:res_sat}
\end{eqnarray}
\end{widetext}
This summation over $k$ values in the Brillouin zone needs to be performed numerically. We have plotted the residual energy from Eq.~(\ref{eq:resi}) as a function of $n$ in Fig.~(\ref{fig:2a}). At large $n\to\infty$, the residual energy saturates to a value given by Eq.~(\ref{eq:res_sat}), indicating a steady state where the system stops absorbing energy. The $n$-independent part of Eq.~(\ref{eq:resi}) represents the steady state. At very high frequencies, the residual energy is zero and does not vary with $n$. This can be understood from the fact that high frequency corresponds to a very fast drive, causing the system to remain in the ground state throughout the driving. In Fig.~(\ref{fig:2b}), we have plotted the residual energy as a function of drive frequency. The dashed line represents the residual energy calculated from the saturation value given by Eq.~(\ref{eq:res_sat}). For large but finite $n$ (e.g., $n=500$ in this case), we observe oscillations around the saturation value.
The saturation value of the residual energy simplifies further for $g_1=0$ and $\omega=2\pi$, and is given by
\begin{eqnarray}
E_{res}^{sat}|_{g_1=0}&=&\frac{1}{L} \sum_{k > 0} \Big((g_0 - \cos k) n_{kx}^2 - n_{kx} n_{kz} \sin k \Big),
\end{eqnarray}
However, by substituting the expressions for $n_{kx}$ and $n_{kz}$ from Eq.~(\ref{eq:delta_para}), the equation above simplifies to zero. This is because when $g_1 = 0$, the Hamiltonian becomes time-independent, which implies that energy is conserved. Additionally, similar expressions for the current operator, given by $\hat{J} = \frac{i}{L} \sum_{i=1}^{L} \left(c_{i}^{\dagger} c_{i+1} - c_{i+1}^{\dagger} c_{i}\right)$, can also be derived. The behavior of the current is similar to that of the residual energy and matches with results found in the literature for different driving protocols\cite{Somnath}.
\section{Magnetization}
\label{sec_dyn_fre}
\begin{figure}[htb]
\begin{center}
\sg[~]
{\includegraphics[width=0.494\columnwidth]{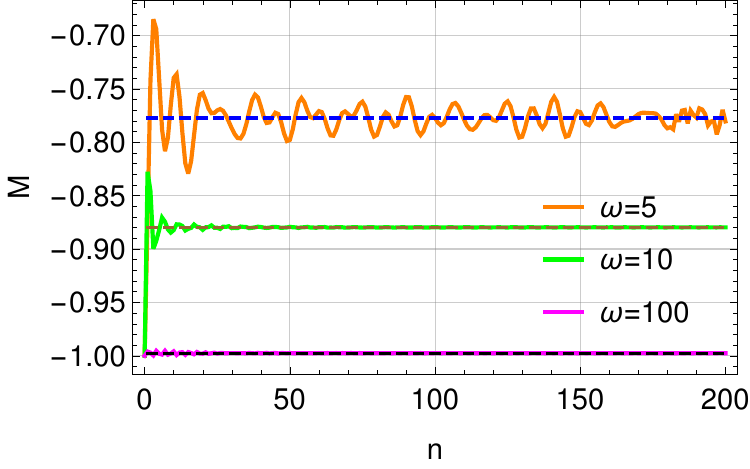}\label{fig:3a}}
\sg[~]
{\includegraphics[width=0.494\columnwidth]{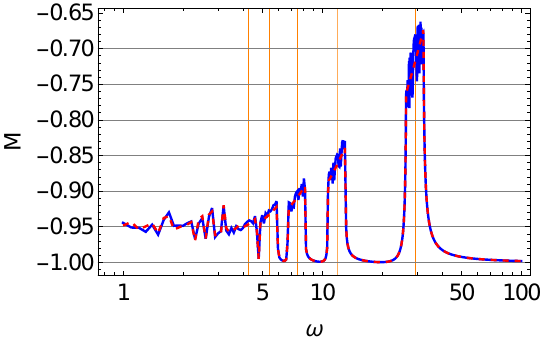}\label{fig:3b}}
\end{center}
\caption{(a) The behavior of magnetization as a function of $n$. (b) Magnetization after $n=500$ drive cycles as a function of drive frequency. The dashed line shows the magnetization computed using the large $n$ saturation value from the $n$-independent part of Eq.~(\ref{eq:mag}). The parameters used in these figures are the same as in Fig. (\ref{fig:2}).}
\label{fig:3}
\end{figure}

In this section, we study the magnetization $M$, defined as the expectation value of $\hat{M} = \sum_{i} \sigma_{i}^{z}$. At stroboscopic intervals, the magnetization is given by
\begin{equation}
M(n)=\frac{2}{L}\sum_{k > 0}\left(2 |u_k(n)|^2-1\right).
\label{eq:mag_def}
\end{equation}
This quantity is analogous to the density of the Kitaev chain. Following the same calculations as prescribed in the previous section, we obtain the expression for the magnetization after $n$ cycles as
\begin{eqnarray}
M(n)&=&\frac{2}{L}\sum_{k > 0}\left(\Gamma_k^2(1-\cos 2n\theta_k)-1\right).
\label{eq:mag}
\end{eqnarray}
We have plotted the expression for magnetization given by Eq.~(\ref{eq:mag}) as a function of $n$ in Fig.~\ref{fig:3a}. The initial state from which the evolution starts is $\left(0,1\right)^{\mathbf{T}}$ for each $k$ mode, which corresponds to the initial magnetization $M(0) = -1$. As discussed in the previous section, at large $n$, the magnetization saturates to a value given by the $n$-independent part of Eq.~(\ref{eq:mag}). In Fig.~\ref{fig:3b}, we plot the magnetization as a function of drive frequency. For the square-pulse protocol, it has been observed that for certain frequencies, the magnetization remains at its initial value and remains almost constant with $n$, leading to a phenomenon known as dynamical freezing. We note that for the periodic kick protocol, this behavior is only present for large values of $g_0$. In Fig.~\ref{fig:3b}, we use $g_0 = 10$. For large $g_0$, $\Gamma_k \rightarrow 0$ for all values of $k$, leading to $M = -1$ except at specific frequencies highlighted by the orange vertical lines in Fig.~\ref{fig:3b}. This behavior is substantially different from the dynamical freezing reported in the literature. 
To understand this, we expand $\Gamma_k$ for large $g_0$. In this limit, the energy dispersion in Eq.~(\ref{eq:dis}) can be approximated as $\epsilon_k \simeq g_0$ and $n_{kz} \simeq 1$. With this approximation, $\Gamma_k$ from Eq.~(\ref{eq:ukn}) simplifies to
\begin{eqnarray}
\Gamma_k &\simeq& \left(\frac{\sin k \sin(Tg_0)}{g_0 \sin(g_1 + Tg_0)}\right)^2.
\label{eq:gamk_a}
\end{eqnarray}
For any arbitrary $k$, all $\sin$ terms are less than 1. Moreover, $g_0^2$ in the denominator makes $\Gamma_k$ almost zero for all $k$, except when the $\sin(g_1 + Tg_0)$ term in the denominator is close to zero for all $k$. This condition translates to $g_1 + Tg_0 = m\pi$, where $m$ is an integer. This gives the frequencies $\omega^{*} = \frac{2g_0\pi}{m\pi - g_1}$, close to which the freezing phenomenon does not occur. In Fig.~\ref{fig:3b}, with $g_0 = 10$ and $g_1 = 1$, we see that close to the frequency $\omega^{*} = \frac{20\pi}{m\pi - 1}$, highlighted by the orange vertical line, the magnetization deviates from -1 and can be obtained by using the limiting value of Eq.(\ref{eq:gamk_a}). For higher values of $m$, the vertical lines become very close to each other, resulting in almost flat values of magnetization slightly away from -1. Finally, we can conclude that for the delta-kick protocol and for the high values of $g_0$, the magnetization is always dynamically frozen except close to some specific frequencies, which we discuss in detail in this section.
\section{fidelity}
\label{sec_fide}
In this section, we present our findings for the fidelity defined as $\chi(n) = \prod_{k} |\langle \psi_k(0) | \psi_k(nT) \rangle|^2$. Using the expression for the wavefunction after the $n$th cycle of the delta-kick protocol given by Eq.~(\ref{eq:kick}), we calculate the logarithm of the fidelity $g(n)$ as follows:
\begin{eqnarray}
g(n) &=& \ln \chi(n)\nonumber=\ln \left(\prod_{k > 0} |\langle \psi_k(0) | \psi_k(nT) \rangle|^2 \right) \nonumber \\
&=& \frac{2}{L} \sum_{k > 0} \ln \left(|v_k(n)|^2 \right) \nonumber=\frac{2}{L} \sum_{k > 0} \ln \left(1 - |u_k(n)|^2 \right) \nonumber \\
&=& \frac{2}{L} \sum_{k > 0} \ln \left|1 - \frac{\Gamma_k^2}{2} (1 - \cos 2n \theta_k) \right|.
\label{eq:fid}
\end{eqnarray}
\begin{figure}[htb]
\begin{center}
\sg[~]
{\includegraphics[width=0.494\columnwidth]{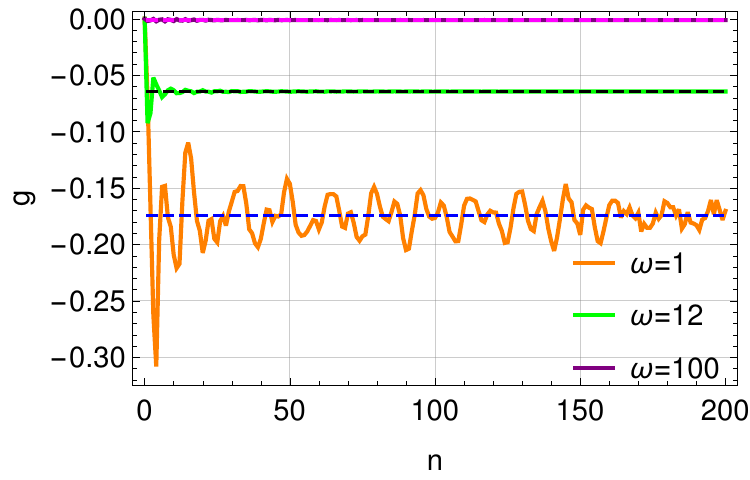}\label{fig:4a}}
\sg[~]
{\includegraphics[width=0.494\columnwidth]{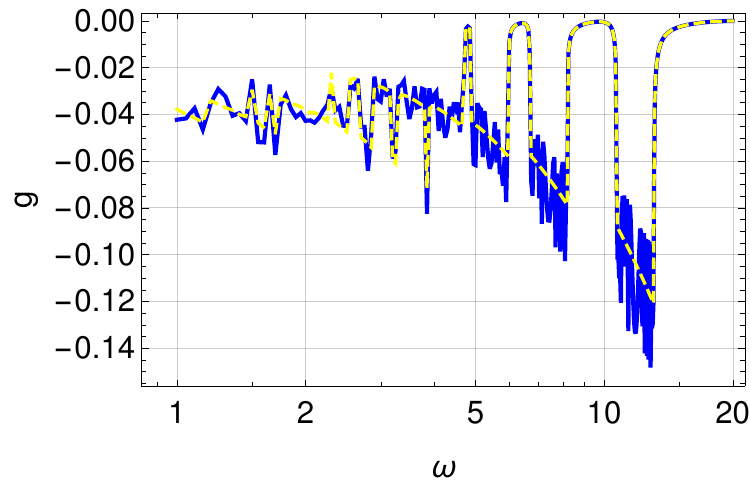}\label{fig:4b}}
\end{center}
\caption{(a) Log fidelity as a function of $n$. The values of the frequencies are given in the inset. (b) The saturation value of the log fidelity after $n=500$ drive cycles is plotted as a function of frequency. The dashed line indicates the $g$ computed using the large $n$ saturation value given by Eq.~(\ref{eq:fid_sat}). The parameters used in these figures are the same as in Fig. (\ref{fig:2}).}
\label{fig:4}
\end{figure}
We have plotted $g(n)$ as a function of $n$ in Fig.~\ref{fig:4a}. Initially, $g(n)$ starts from a certain value, decays, and eventually saturates at a value for large and finite $n$. Note that the highly oscillating term is inside the logarithm, and in this case will contribute to the integral in Eq.~(\ref{eq:fid}) even for large but finite $n$. One can show that all the even power terms in the expansion of the logarithm are nonzero. Following the prescription discussed by S. Sharma et al. in Ref.~\cite{Shraddha}, the saturation value is found to be
\begin{eqnarray}
g_{sat} &=& -\frac{4}{L} \sum_{k > 0} \ln \left(\frac{2}{1 + \sqrt{1 - \Gamma_k^2}} \right).
\label{eq:fid_sat}
\end{eqnarray}
In Fig.~\ref{fig:4b}, we plot $g(500)$ as a function of frequency. The dashed line indicates the $g(n \to \infty)$ value given by Eq.~(\ref{eq:fid_sat}). We observe that $g(500)$ oscillates around the $g(n \to\infty)$ value, and for particular frequencies where $\Gamma_k=0$, $g(n)$ approaches 1, similarly to the behavior of magnetization discussed earlier.
\section{Correlation function}
\label{sec:corr}
\begin{figure}[htb]
\begin{center}
\sg[~]{\includegraphics[width=0.494\columnwidth]{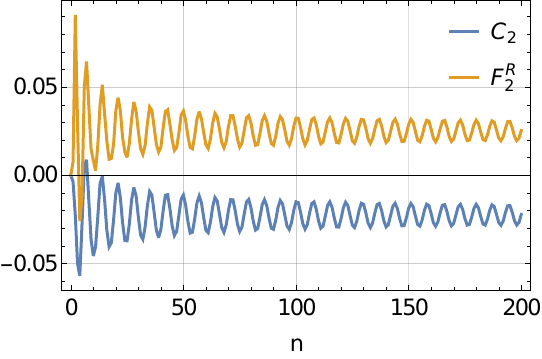}\label{fig:5a}}
\sg[~]{\includegraphics[width=0.494\columnwidth]{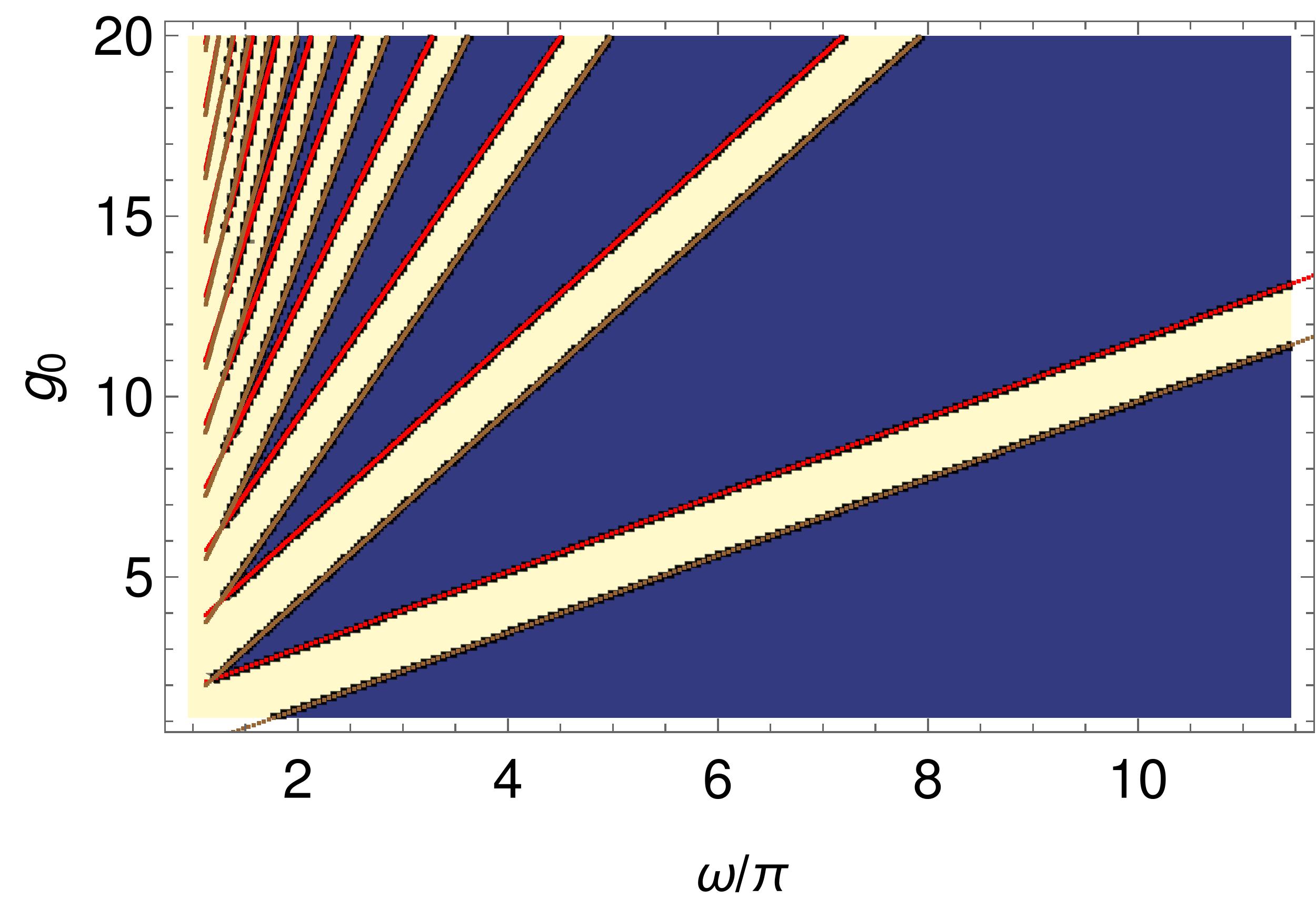}\label{fig:5b}}
\end{center}
\caption{(a) The correlation function $C_{2}$ and $F_2^{R}$ is plotted as a function of $n$ using Eq.(\ref{eq:cana}, \ref{eq:fanar}). Here we have used $g_0=2$, $g_1=1$ and $\omega=5$.(b)The phase diagram of dynamical transition in relaxation behavior. The blue region signifies relaxation with power law $-\frac{3}{2}$ while the yellow region has power-law $-\frac{1}{2}$. At the transition point red and brown lines (the solution of Eq.(\ref{eq:wc})) the power-law of the relaxation behavior is $-\frac{3}{4}$. Here we have taken $g_1=1$.}
\end{figure}

\begin{figure}[htb]
\begin{center}
\sg[~]{\ig[width=0.494\columnwidth]{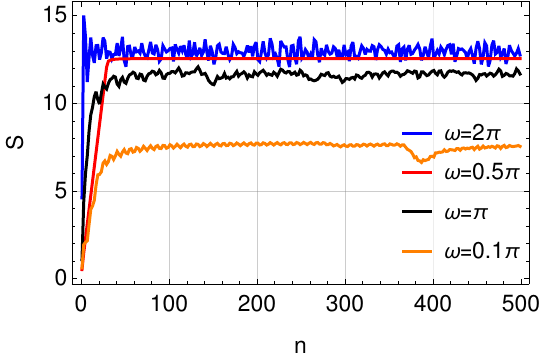}\label{fig:7a}}
\sg[~]{\ig[width=0.494\columnwidth]{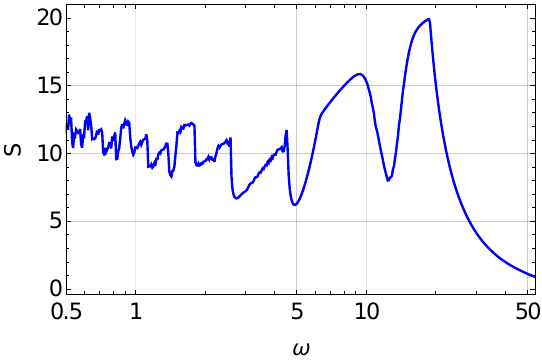}\label{fig:7b}}
\end{center}
\caption{(a) The behavior of entanglement entropy as a function of drive cycle $n$. The values of the frequencies are given in the inset. We have taken $g_0=2$, $g_1=1$ in this figure. (b) The saturation value of entanglement entropy after $n=200$ drive cycles as a function of drive frequency. Here $g_0=2$, $g_1=1$, $L = 1000$ and $l = 50$.}
\label{fig:7}
\end{figure}

In this section, we utilize the analytical expression for $|\psi(nT)\rangle$ provided in Eq.~(\ref{eq:nth}) to investigate the dynamical relaxation behavior of our system under delta-kick driving. We begin by calculating two nontrivial fermionic correlators to examine the dynamical relaxation of the system.
\cite{Krishentangle,Arnab_Sen,Diptiman,Makki,Madhumita}. The fermionic correlators are defined as:
\begin{eqnarray}
C_{ij}(t) &=& \frac{2}{L} \sum_{k > 0} |u_{k}(t)|^2 \cos(k(i-j)), \\
F_{ij}(t) &=& \frac{2}{L} \sum_{k > 0} u_{k}^{*}(t) v_{k}(t) \sin(k(i-j)).
\label{eq:cor}
\end{eqnarray}
Understanding the behavior of the correlation functions is crucial as it contains the information about quantum criticality. Recent studies have identified that singularities in certain out-of-time-ordered correlation functions for finite-length products indicate the presence of dynamical phase transitions\cite{Amit2}. Using Wick's theorem, all correlation functions of the system can be expressed in terms of these two nontrivial correlations. As a result, we proceed to determine the $C$ and $F$ correlation matrices stroboscopically. It is evident from the definitions that for $i=j$, the diagonal terms of the $C$ matrix are all equal, whereas the diagonal terms of the $F$ matrix are all zero. Using the fact that the elements of the $C$ and $F$ matrices depend only on $i-j$, we define an integer variable $m=i-j$ and obtain the analytical expression for all the nonzero terms of the $C$ matrix:
\begin{eqnarray}
C_{m}(n) &=& \frac{1}{L} \sum_{k > 0} \Gamma_k^2 \cos(mk) (1 - \cos 2n \theta_k).
\label{eq:cana}
\end{eqnarray}
The $F$ matrix is complex, and following a similar prescription, we also obtain the analytical expression for the $F$ matrix:
\begin{eqnarray}
F_{m}^{R}(n)&=&-\frac{1}{L}\sum_{k>0}\frac{\Gamma_k\sin mk}{\sin\theta_k} (\sin g_1\cos\Phi_k\nonumber\\
&&+n_{kz}\cos g_1\sin\Phi_k)(1-\cos 2n\theta_k),\label{eq:fanar} \\
F_{m}^{I}(n)&=&\frac{1}{L}\sum_{k>0}\Gamma_k\sin mk\sin 2n\theta_k.
\label{eq:fanai}
\end{eqnarray}
Here $F_{m}^{R}$ and $F_{m}^{I}$ are the real and imaginary parts of the complex $F$ matrix, respectively. From Eqs.~(\ref{eq:cana}, \ref{eq:fanar}, \ref{eq:fanai}), it is clear that the terms independent of $n$ survive in the limit of large but finite $n$, leading to convergence to a generalized Gibbs ensemble (GGE)\cite{Armab_Das2,Abhisek}. At large $n$, the imaginary part of the $F$ matrix vanishes, and both $C$ and $F$ matrices become real. This can be explained by the fact that the system absorbs energy from the periodic driving and heats up to a steady state characterized by the conserved quantities of the integrable system. We have plotted the correlation functions as a function of $n$ in Fig.\ref{fig:5a} for $m=2$.\\
Next, we study the approach to the steady state. We have plotted in Fig.~(\ref{fig:6}) the absolute value of the $n$-dependent part of the correlation matrix, $\delta C_m(n)=\frac{1}{L}\sum_{k > 0}\Gamma_k^2\cos(mk)(\cos 2n\theta_k)$, which decays following a power law as reported in earlier literature\cite{Krishentangle,Diptiman}. This power-law behavior arises from both slowly varying and highly oscillating terms and is universal in all local observables. Using the saddle-point approximation discussed in Ref.~\cite{Krishentangle}, one can approximate the $n$-dependent part of the correlation function at large $n$. Since the $n$-dependent part is highly oscillatory, most of the contribution to the integral in Eqs.~(\ref{eq:cana}, \ref{eq:fanar}, \ref{eq:fanai}) comes from the saddle point at $k = k_0$, where $\theta_k$ is at an extremum. The value of $k_0$ can be calculated by differentiating Eq.~(\ref{eq:thetk}) with respect to $k$ and equating it to zero. Although an analytical expression for the first derivative of $\theta_k$ is possible, solving the resulting transcendental equation numerically yields the value of $k_0$. In the delta-kick protocol, the extremum of $\theta_k$ usually occurs at $k_0 = 0$ or $\pi$, and at these points, $\Gamma_k = 0$ due to the presence of $\sin k$ in the numerator from $n_{kx}$. This causes the slowly varying term to vanish at $k = k_0$, leading to a power law behavior characterized by an exponent of $-3/2$. Conversely, for some values of $\omega$, the saddle point occurs at $k_0 \ne 0$ or $\pi$, where the slowly varying term is non-zero, resulting in a power law behavior characterized by an exponent of $-1/2$. A dynamical transition between these two power law behaviors occurs as a function of drive frequency, with the critical frequency $\omega_c$ defining the transition. In Fig.~(\ref{fig:5a}), we reproduce the phase diagram of this dynamical transition, first reported in Ref.~\cite{Krishentangle}. In the blue region of the phase diagram, the roots of $\frac{d\theta_k}{dk} = 0$ are only $0$ and $\pi$, indicating a $-3/2$ relaxation behavior. In the yellow region, other roots of the equation $\frac{d\theta_k}{dk} = 0$ are possible, leading to a $-1/2$ relaxation behavior. At the transition from the blue to the yellow region, an extra root appears in the equation $\frac{d\theta_k}{dk} = 0$, implying that $\frac{d^2\theta_k}{dk^2} = 0$ exactly at the transition point. This gives the critical frequency $\omega_c$ where the transition occurs, defined by the solution to the equation:
\begin{eqnarray}
\tan \left(\frac{z_{\pm}}{\omega}\right) = \mp \frac{\left(\frac{z_{\pm}g_0}{\omega}\right) \tan g_1}{\tan g_1 \pm \left(\frac{z_{\pm}g_0}{\omega}\right)},
\label{eq:wc}
\end{eqnarray}
where $z_{\pm} = 2\pi(g_0 \pm 1)$. The $+$ and $-$ signs correspond to $k_0 = \pi$ and $0$, respectively, reflecting the fact that the extra root of the equation $\frac{d\theta_k}{dk} = 0$ arises either from the $0$ or $\pi$ side. We numerically solve Eq.~(\ref{eq:wc}) and plot it in Fig.~(\ref{fig:5a}). The red and brown lines correspond to the solutions for the $+$ and $-$ signs of the equation, respectively. This Eq.(\ref{eq:wc}) is a new result and matches with Ref.~\cite{Krishentangle}.
\begin{widetext}
\begin{center}
\begin{figure}[htb]
\begin{center}
\sg[~]
{\includegraphics[width=5.7cm]{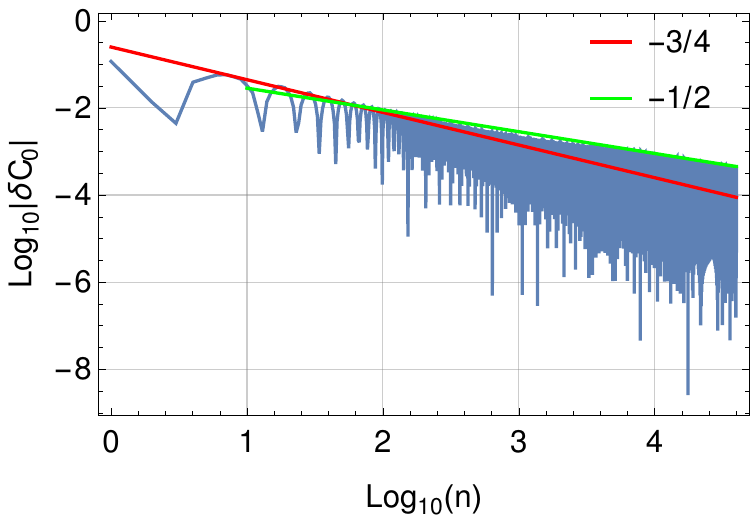}\label{fig:6a}}
\sg[~]
{\includegraphics[width=5.7cm]{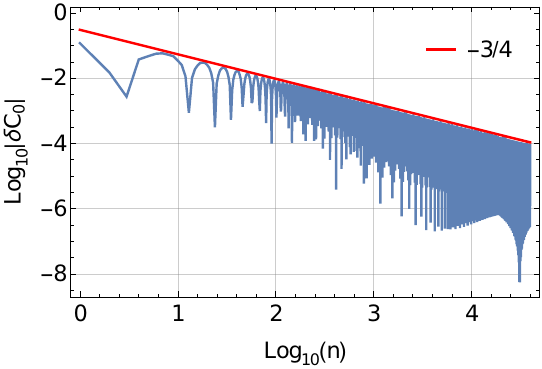}\label{fig:6b}}
\sg[~]
{\includegraphics[width=5.7cm]{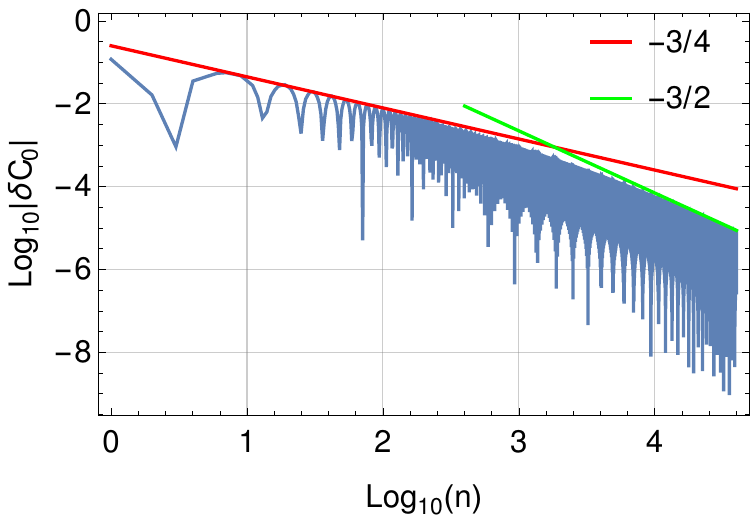}\label{fig:6c}}
\end{center}
\caption{We have plotted $\log|\delta C_0(n)|$ as a function of $\log n$. (a) $\omega = 8.24094$, slightly below the critical frequency. (b) $\omega_c = 8.26094$, at the critical frequency. (c) $\omega = 8.28094$, slightly above the critical frequency. The red and green lines are guides to the eye. Here we have taken $g_0=2$, $g_1=1$ and system size $L=80000$.}
\label{fig:6}
\end{figure}
\end{center}
\end{widetext}
Recently, it has been observed in~\cite{Diptiman} that at the dynamical transition point, the relaxation behavior of the $n$-dependent part of the correlators shows a different power law behavior characterized by an exponent of $-3/4$. At this transition point, not only are the first and second derivatives of $\theta_k$ zero, but the third derivative is also zero. The first non-zero term in the Taylor series expansion of $\theta_k$ is of fourth order. Moreover, on the transition line, $k_0 = 0$ or $\pi$, which corresponds to $\Gamma_k = 0$. Therefore, we must modify the saddle-point approximation defined in Ref.~\cite{Krishentangle} by including higher-order terms. The modified approximation is given by\cite{Book1,Book2}:
\begin{eqnarray}
\int dk \, f(k) e^{in \theta_k} \simeq e^{-in \theta_{k_0}} e^{i \frac{7 \pi}{8}} \frac{6 f''(k_0) \Gamma\left(\frac{3}{4}\right)}{(24)^{\frac{1}{4}} (|\theta_{k_0}''''|)^{\frac{3}{4}}} n^{-\frac{3}{4}},
\end{eqnarray}
where $f(k)$ is the slowly varying part of the correlators that contains $\Gamma_k$, $f(k_0)=0$ and $f'(k_0)=0$. The $n^{-\frac{3}{4}}$ term clearly explains the critical relaxation behavior.
In Fig.~(\ref{fig:6}), we have plotted the $|\delta C_{0}(n)|$ as a function of $n$ for three different frequencies. The Fig.~(\ref{fig:6a}) shows the relaxation behavior below the critical frequency but very close to it. Due to this, the initial relaxation follows a power law with an exponent of $-3/4$, but at large $n$, a transition to an exponent of $-1/2$ is observed. Fig.~(\ref{fig:6b}) shows critical relaxation characterized by the exponent $-3/4$. The Fig.~(\ref{fig:6c}) shows dynamical relaxation above the critical frequency but very close to it, where a transition from $-3/4$ to $-3/2$ is visible. Using the saddle-point approximation at the critical point, we find a simplified numerical expression for $|\delta C_{0}(n)|$ that demonstrates the power-law scaling:
\begin{eqnarray}
|\delta C_{0}(n)| \simeq 1.79067 \frac{\cos(3.00142 n)}{n^{\frac{3}{4}}}.
\end{eqnarray}
Similar expressions for the $\delta F_{m}^{R}(n)$ and $\delta F_{m}^{I}(n)$ can also be derived and are straightforward to calculate.\\
Using the analytical expressions for the $C$ and $F$ matrices given by Eqs.~(\ref{eq:cana}, \ref{eq:fanar}, \ref{eq:fanai}), we compute the correlator matrix $\Pi$ defined as:
\begin{eqnarray}
\Pi_{n}(l) = \left(
\begin{array}{cc}
{\mathbb I}_l - C & F \\
F^{*} & C
\end{array}
\right),
\end{eqnarray}
where $l$ is the subsystem size and the indices $i, j$ of the $C$ and $F$ matrices are restricted to within the subsystem. The entanglement entropy\cite{Kitaev,entangrev1,entangrev2,cardy1} is given by $S_{n}(l) = -\sum_{i} \rho_{i} \ln \rho_i$, where $\rho_{i}$ are the eigenvalues of the $\Pi_{n}(l)$ matrix. We have numerically diagonalized the correlator matrix to compute the entanglement entropy, which is plotted in the Fig.~(\ref{fig:7}). Our analytical expressions simplify the calculation, as numerical computation of the $n$-th power of the evolution matrix is not required. 
In Fig.~(\ref{fig:7a}), we plot the entanglement entropy as a function of $n$. It starts from zero since the initial state is a product state and saturates to a value proportional to the length\cite{hasting1} of the subsystem at large $n$. Fig.~(\ref{fig:7b}) shows the behavior of the entanglement entropy as a function of drive frequency after $n=200$ drive cycles. At large drive frequencies, the system fails to respond to the external periodic driving, and the system remains in the initial product state, resulting in the entanglement entropy remaining at zero even after many drive cycles.
\section{Sinusoidal Driving}
\label{sec:sinu}
In this section, we analyze the TFI model driven by a sinusoidal periodic protocol.  In contrast to the delta-kick pulse, the unitary time evolution operator for the one complete drive cycle cannot be derived analytically, in this case. The protocol we use here is given by  
\begin{equation}  
g(t)=g_{av} + \delta g\cos(\omega t),  
\end{equation}  
where $g_{av}=(g_0+g_1)/2$ is the average value of the driving parameter and $\delta g=(g_0-g_1)/2$ represent the amplitude of the periodic driving. $\omega$ is the drive frequency. Since the Hamiltonian at different times does not commute for sinusoidal protocol, we use numerical methods to obtain the time dependent state. The state of the system at any given time can be expressed as $|\psi_k(t)\rangle = u_k(t)|0\rangle + v_k(t)|1\rangle$, where $|0\rangle$ and $|1\rangle$ represent the diabatic basis states. The time-dependent Schr\"odinger eq in terms of $u_k(t)$ and $v_k(t)$ simplifies to 
\begin{equation}
i\frac{d}{dt}
\begin{pmatrix}
u_k(t) \\
v_k(t)
\end{pmatrix} =
\begin{pmatrix}
g(t) - \cos k & \sin k \\
\sin k & -(g(t) - \cos k)
\end{pmatrix}
\begin{pmatrix}
u_k(t) \\
v_k(t)
\end{pmatrix}.
\label{eq:sch}
\end{equation}
Solving the above set of equations numerically, we get the time-dependent state after $n$th drive cycles represented as $(u_k(n)~v_k(n))^{T}$. The ground state wavefunction at $t = 0$ serves as the initial condition. However, solving the above set of differential equation for longer time period, i.e. large $n$, would accumulate errors inherent in any numerical method. To address this issue, we solve the equation above for a single time period. By utilizing the time-dependent state after one time period, the evolution operator for that duration has been constructed. The general form of evolution operator for one period is given by Eq.~(\ref{eq:uk}) and the components of the unitary evolution matrix $U_k(T)$ can be expressed in terms of both the initial and final state wavefunctions, and written as:
\begin{eqnarray}
\alpha_k&=&u_k^*(0) u_k(1) + v_k(0) v_k^*(1)\\
\beta_k &=& u_k^*(0) v_k(1) - v_k(0) u_k^*(1).
\end{eqnarray}
Once the evolution operator for a single time period is obtained, the time-dependent state after any finite $n$ drive periods can be determined using the procedure described in Eq.~(\ref{eq:nth}). Utilizing the solution of the time-dependent state after the $n$th drive cycle, we have numerically computed the expectation value of the same observables as those computed earlier. 
\begin{figure}[htb]
\begin{center}
\sg[~]{\ig[width=0.494\columnwidth]{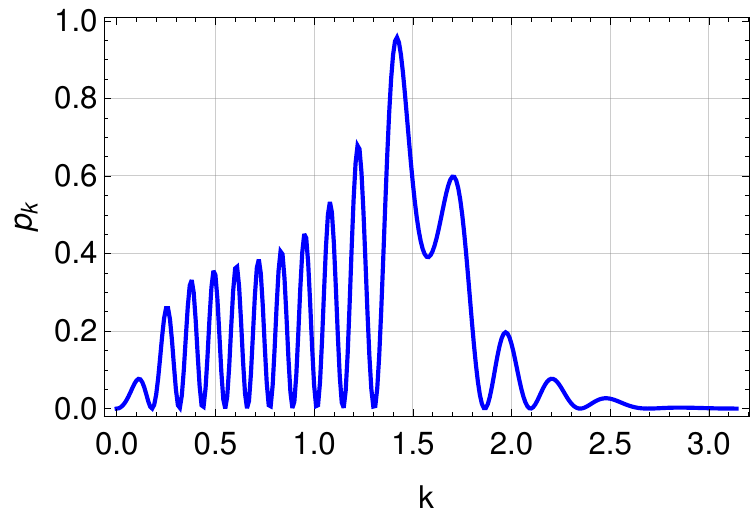}\label{fig:8a}}
\sg[~]{\ig[width=0.494\columnwidth]{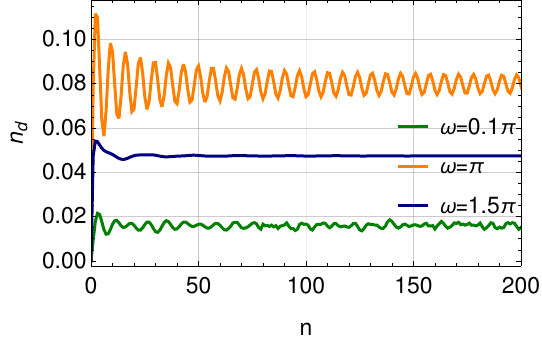}\label{fig:8b}}
\sg[~]{\ig[width=0.494\columnwidth]{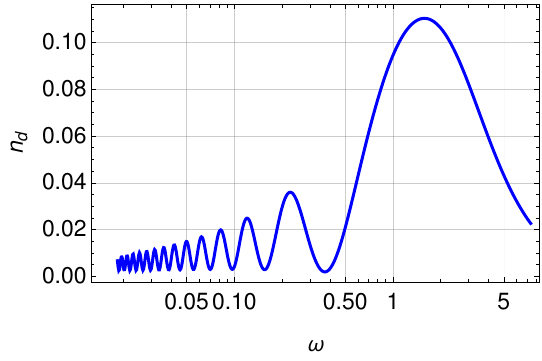}\label{fig:8c}}
\sg[~]{\ig[width=0.494\columnwidth]{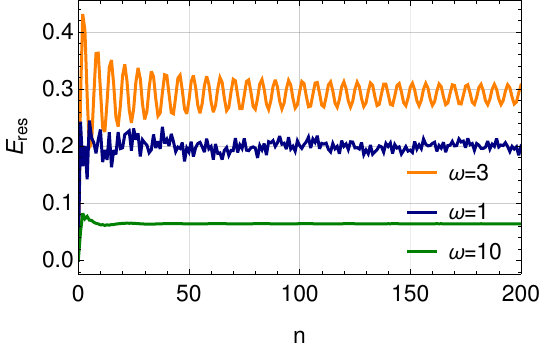}\label{fig:8d}}
\end{center}
\caption{(a) The behavior of $p_k$ as a function of $k$ for $n=15$, plotted as a solid blue line for sinusoidal pulse protocol. Higher $n$ corresponds to more oscillations. (b) The defect density generated after $n$ cycles is plotted against drive cycle $n$ for different frequencies. (c) The behavior of $n_d$ as a function of $\omega$ after one drive cycle is plotted as a solid blue line for the sinusoidal pulse protocol. (d) The behavior of $E_{\text{res}}$ as a function of $n$ for different angular frequencies $\omega$ is plotted for the sinusoidal pulse protocol. The parameters used are: $L = 2048$, $g_0 = 2$, and $g_1 = 0.5$.}
\label{fig:8}
\end{figure}
\subsection{Results}
In this subsection, we present the results obtained for the TFI model driven by a sinusoidal protocol. At $t=0$, the value of the transverse field is $g=g_0$ and the TFI Hamiltonian is diagonalized for $g_0$ to find the ground state $|\psi_{k}^{g}(0)\rangle$ and excited state $|\psi_{k}^{ex}(0)\rangle$. Using this ground state as the initial state, we numerically determine the time-dependent state. For a periodic driving, the system returns to the value $g_0$ after each $n$th drive cycle, and we compute the overlap with the excited state to determine the excitation probability $p_k(n)=|\langle\psi_{k}(nT)|\psi_{k}^{ex}(0)\rangle|^2$. We have plotted the excitation probability as a function of $k$ in Fig.~\ref{fig:8a}. Next, we have summed over the excitation probability over all the k modes and divide by number of site to calculate the defect density $n_d$. We have computed the defect density after each drive cycle and plotted as a function of n in Fig.~\ref{fig:8b}. In Fig.~\ref{fig:8c} we have plotted the defect density after one drive cycle as a function of frequency. Finally, in Fig.~\ref{fig:8d} residual energy is plotted with $n$. 
%
\begin{figure}[htb]
\begin{center}
\sg[~]{\ig[width=0.494\columnwidth]{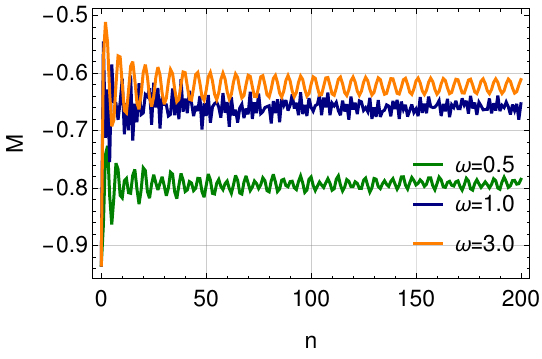}\label{sin_mag}}
\sg[~]{\ig[width=0.494\columnwidth]{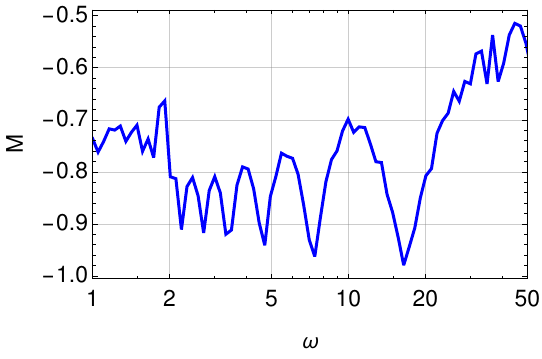}\label{sin_mag_w}}\\
\sg[~]{\ig[width=0.494\columnwidth]{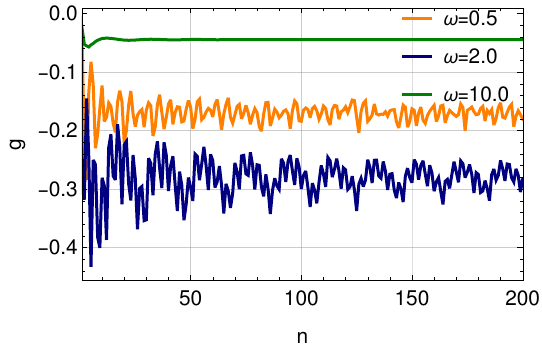}\label{sin_fid}}
\sg[~]{\ig[width=0.494\columnwidth]{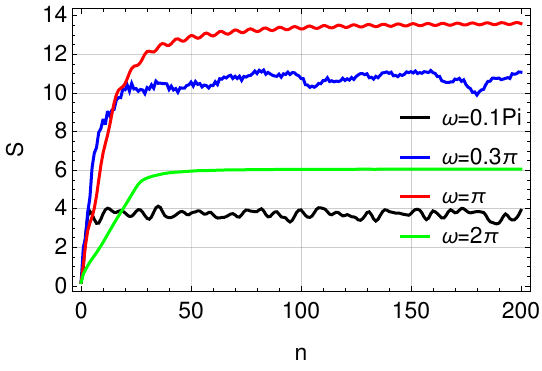}\label{sin_ent}}
\end{center}
\caption{(a) The behavior of $M$ as a function of $n$ for different angular frequencies $\omega$ is plotted for the sinusoidal pulse protocol. (b) Magnetization after $n=1000$ drive cycles as a function of drive frequency $\omega$ plotted in solid blue line for a sinusoidal pulse. For this figure only we have taken $g_0=-g_1=20$. (c) The behavior of $g$ as a function of $n$ for different drive frequencies $\omega$ is plotted for sinusoidal pulse protocol. The parameters used are: $L = 2048$, $g_0 = 2$, and $g_1 = 0.5$.}
\label{fig:11}
\end{figure}

In Fig.~(\ref{sin_mag}), we show the magnetization as a function of the drive cycle. In contrast, Fig.~(\ref{sin_mag_w}) presents the magnetization as a function of the drive frequency after $n = 1000$ drive cycles. For sinusoidal driving, the magnetization also saturates at large $n$, although at certain frequencies, a freezing phenomenon is observed. The logarithmic fidelity is displayed in Fig.~(\ref{sin_fid}), and the entanglement entropy is shown in Fig.~(\ref{sin_ent}). All of these figures exhibit a similar behavior to the delta kick protocol.
\section{Discussion}
\label{sec_diss}
In this paper, we have studied periodically driven integrable systems, focusing specifically on the long-time dynamics of a periodically delta-kicked transverse field Ising model. We derive an analytical expression for the time-evolution operator over one period and use this to obtain the exact analytical expression for the time-dependent state at stroboscopic intervals. This time-dependent wave function enables the calculation of the expectation value of any observable after the $n$-th drive cycle.
Initially, we derive the expression for the defect density $n_d$, which consists of two components: a highly oscillatory part that varies with $\theta_k$, the eigenvalue of the evolution operator over one period. The number of oscillations in the excitation probability for each $k$ increases with drive cycle $n$. The second term in $n_d$ is independent of $n$, representing the periodic-GGE and providing the saturation value for large $n$. This feature is universal across the expectation values of all observables.
Next, we compute the residual energy, magnetization, and log fidelity. For large values of $g_0$, the magnetization exhibits dynamic freezing for almost all drive frequencies, except for a few specific frequencies where it deviates from its initial value. These frequencies, given by $\omega^* = \frac{2g_0 \pi}{m \pi - g_1}$, are analyzed for their significance. The residual energy and fidelity show similar behavior at these frequencies.
We then derive the analytical expressions for all elements of the correlation matrix after the $n$-th drive cycle and discuss the relaxation behavior of the correlators, which is true for all observables. Depending on the drive frequency, the relaxation behavior changes the slope of the approach to the steady state from $-3/2$ to $-1/2$. This represents a rare example of a dynamical transition and the emergence of universality in driven systems. At the transition point, the critical relaxation is characterized by a slope of $-3/4$. We derive equations for the critical frequencies, assuming $g_0$ and $g_1$ are fixed, and solve these equations numerically, finding the values of critical frequencies which match those reported in the literature.
Before conclusion, we numerically compute the entanglement entropy (EE) using the analytical expressions for the correlation matrix and discuss its properties. Our results for EE are consistent with earlier numerical findings obtained using Floquet theory.
Finally, we apply our formalism to the sinusoidally driven TFI model. We numerically compute the time evolution operator over one drive cycle by solving the Schrödinger equation. Subsequently, we use our formalism to compute the same observables discussed earlier and analyze their properties within the context of the sinusoidal driving.\\
In conclusion, we have explored the long-time dynamics of a periodically driven integrable system in one dimension. Exact analytical expressions for the stroboscopic time-dependent wave function after the $n$-th drive period have been derived and used this to calculate the defect density, residual energy, magnetization, log fidelity, and two non-trivial quadratic correlators. We have discussed their properties based on these analytical expressions and derived an expression for the dynamical phase boundary that separates different relaxation behaviors of the observables. We have extended our work to the square pulse drive protocol and the sinusoidal drive protocol. Our work opens several avenues for future research, including investigating analytical solutions for other periodic driving protocols and extending the current study to integrable long-range\cite{A.D4} and non-Hermitian systems\cite{Tista1,Tista2,Tista3,nhexp1}.\\
\section{Acknowledgments}
\label{sec_ack}
The authors thank Prof. Diptiman Sen and Prof. Bhabani Prasad Mandal for numerous discussions on related topics.  A.D. acknowledges financial support from the SERB-SRG grant SRG/2022/001145 and the IoE seed grant from BHU IoE/Seed Grant II/2021-22/39963.\\
\appendix
\section{$n$th power of evolution operator}
\label{sec_app}
In this section, we outline the procedure to derive the expression for the $n$th power of the evolution matrix over one period. We extend the calculation developed for a locally periodic system by D. J. Griffiths et al. \cite{Griffiths,Tista4,CFT1,CFT2} to a periodically driven system. We start with the characteristic equation of $U_k(T)$, given by $\det(U_k(T) - \lambda \mathbb{I}_{2\times2}) = 0$, where $\lambda$ represents the eigenvalues of the unitary matrix. Assuming $\xi_k=\frac{1}{2}\Tr(U_k(T))$, we can rewrite the characteristic equation as
\begin{equation}
\lambda^2 - 2\lambda \xi_k + 1 = 0.
\end{equation}
Here, we use the fact that $\det(U_k(T)) = 1$. Applying the Cayley-Hamilton theorem, we obtain
\begin{equation}
U_k^2(T) - 2\xi_k U_k(T) + \mathbb{I}_{2\times2} = 0.
\end{equation}
This equation shows that all higher power of $U_k(T)$ can be expressed as a linear combination of $U_k(T)$ and the identity matrix. We assume:
\begin{equation}
U_k^n(T) = \mathcal{F}_{n-1}(\xi_k) U_k(T) + \mathcal{F}_{n-2}(\xi_k) \mathbb{I}_{2\times2},
\end{equation}
where the coefficients $\mathcal{F}_{n-1}(\xi_k)$ represents polynomials of degree $n-1$ in $\xi_k$, yet to be determined. Using the method of induction, one can straightforwardly derive the recurrence relation for these polynomials, which turns out to be consistent with the recurrence relation of the Chebyshev polynomials of the second kind.  
\section{Square-pulse periodic driving}
\label{sec_app_sq}
\begin{figure}[htb]
\begin{center}
\sg[~]{\includegraphics[width=0.494\columnwidth]{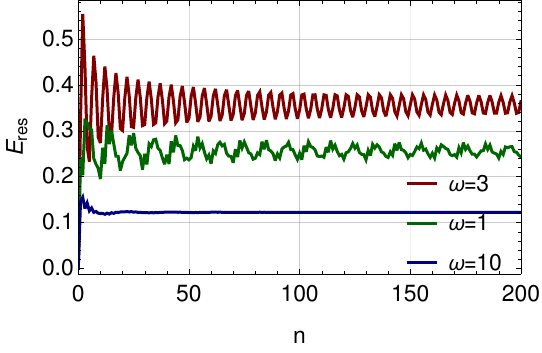}\label{fig:14a}}
\sg[~]{\includegraphics[width=0.494\columnwidth]{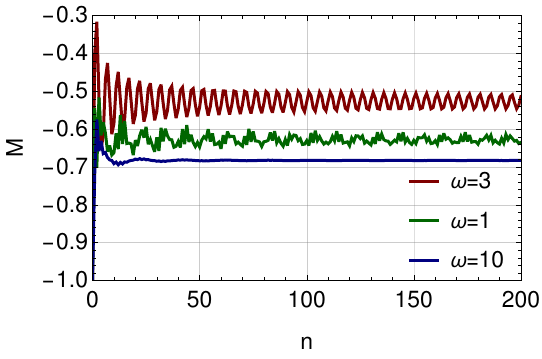}\label{fig:14b}}
\sg[~]{\includegraphics[width=0.494\columnwidth]{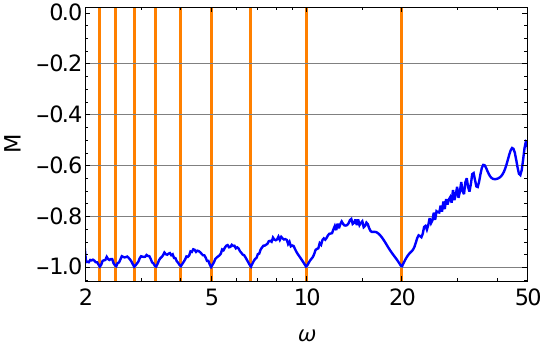}\label{fig:14c}}
\sg[~]{\includegraphics[width=0.494\columnwidth]{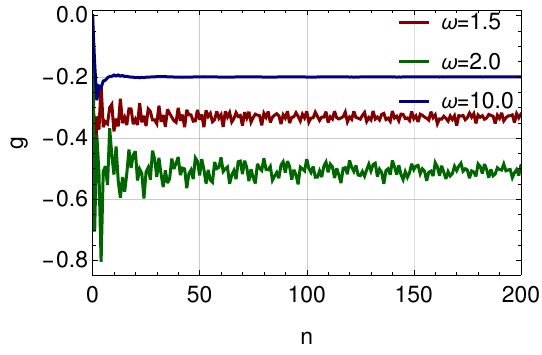}\label{fig:14d}}
\end{center}
\caption{(a) The behavior of $E_{res}$ as a function of $n$ the square pulse.(b) The behavior of $M$ as a function of $n$ for square pulse. (c)Magnetization after $n=1000$ drive cycles as a function of drive frequency $\omega$ plotted in solid blue line for square pulse demonstrating dynamical many-body freezing. Here $g_0=-g_1=20$ is taken only for this figure to highlight the dynamical freezing. The freezing frequencies are highlighted in orange vertical line(d) The behavior of $g$ as a function of $n$ for square pulse. Parameters used: $L=2048$, $g_0=2$, $g_1=0.5$ except figure (c)}
\label{fig:A2}
\end{figure}

\begin{figure}[t]
\begin{center}
\sg[~]{\includegraphics[width=0.449\columnwidth]{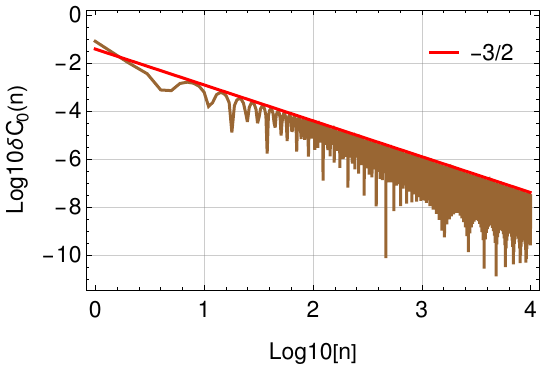}\label{fig:13a}}
\sg[~]{\includegraphics[width=0.449\columnwidth]{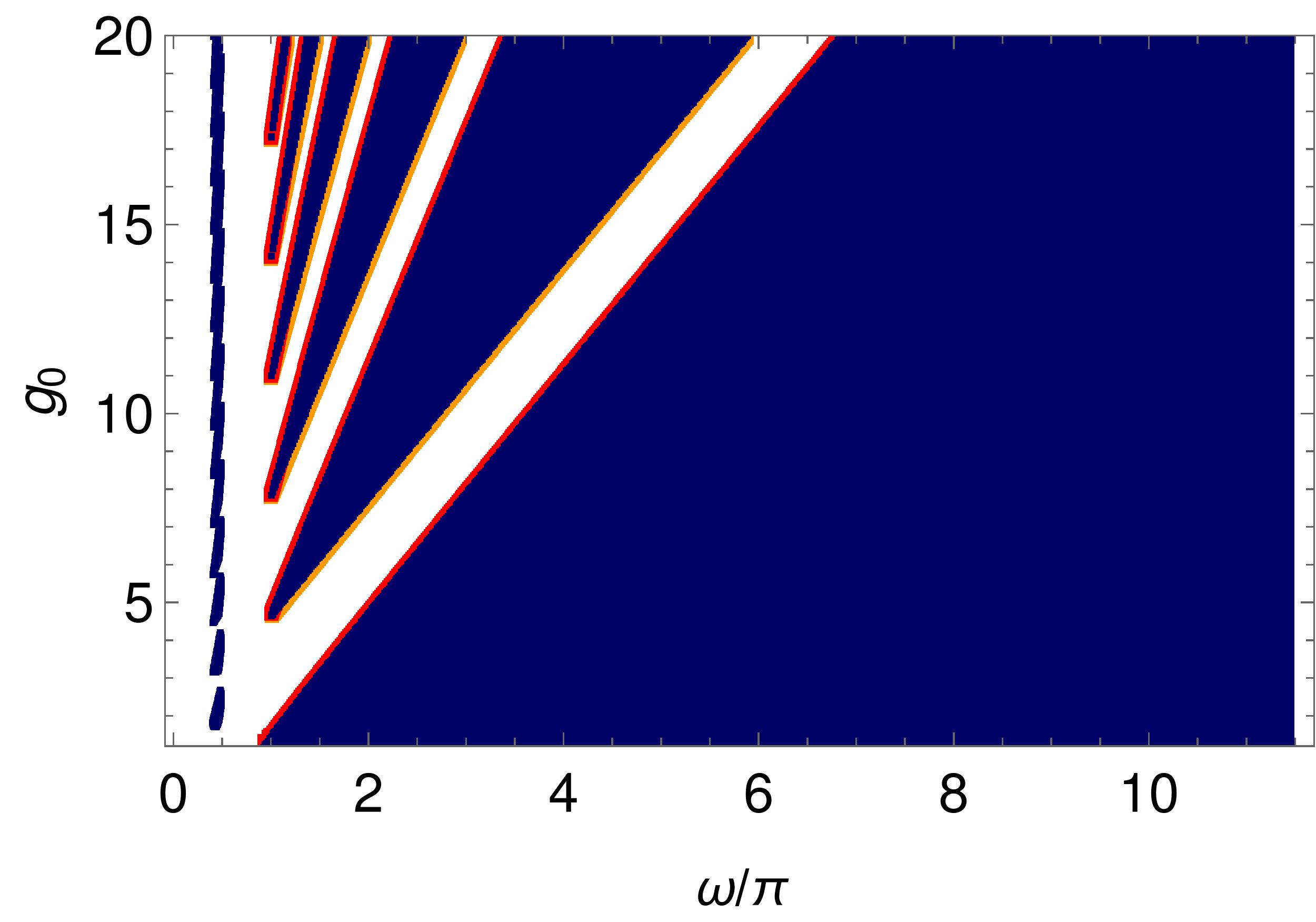}\label{fig:13b}}
\end{center}
\caption{(a) The $\log|\delta C_{0}|$ is plotted as a function of $\log n$ for $\omega=5\pi, g_0=2, g_1=0, L=20000$. (b)The phase diagram of dynamical transition in relaxation behavior. The blue region signifies relaxation with power law $-\frac{3}{2}$ while the white region has powerlaw $-\frac{1}{2}$ in the decay of $\delta C_0(n)$. Here we have taken $g_1=0$. }
\label{fig:13}
\end{figure}

In this section, we discuss the analytical results of the periodically driven TFI model with square-pulse protocol following the same procedure as discussed in the main text. We vary the transverse field $g$ periodically in the following way:
\begin{eqnarray}
g(t)&=& g_0 \quad \text{for } t \leq \frac{T}{2} \nonumber\\
&=& g_1 \quad \text{for } t > \frac{T}{2}. 
\end{eqnarray}
In this case, the unitary time-evolution operator for one drive cycle can be written as:
\begin{eqnarray}
U_k(T)&=& e^{-i H_0 \frac{T}{2}}.e^{-i H_1 \frac{T}{2}},      
\end{eqnarray}
where,
\begin{eqnarray}
H_0&=&(g_0 - \cos k) \tau_3 + (\sin k)\tau_1\\
H_1&=&(g_1 - \cos k) \tau_3 + (\sin k)\tau_1.   
\end{eqnarray}
The general form of the evolution matrix $U_k(T)$ is given by Eq.~(\ref{eq:uk}) and the elements for the square-pulse protocol can be expressed as:
\begin{eqnarray}
\alpha_k &=&\Big( \left(\cos \phi_k^0 - i n_{kz}^0 \sin \phi_k^0 \right)
\left( \cos \phi_k^1 - i n_{kz}^1 \sin \phi_k^1 \right)\nonumber\\
&&-n_{kx}^0 n_{kx}^1 \sin \phi_k^0 \sin \phi_k^1\Big) \\
\beta_k &=& -i \Big(n_{kx}^0 \sin \phi_k^0 \left( \cos \phi_k^1 - i n_{kz}^1 \sin \phi_k^1 \right)\nonumber\\ 
&&+n_{kx}^1 \sin \phi_k^1 \left( \cos \phi_k^0 + i n_{kz}^0 \sin \phi_k^0 \right) \Big)
\end{eqnarray}
Here for the square-pulse protocol the parameters are also modified and given by:
\begin{eqnarray}
\epsilon_k^{0(1)} &=& \sqrt{(g_{0(1)} - \cos k)^2 + \sin^2 k} \\
\phi_k^{0(1)} &=& \frac{T\epsilon_k^{0(1)}}{2} \\
n_{kx}^{0(1)} &=& \frac{\sin k}{\epsilon_k^{0(1)}};~~~n_{kz}^{0(1)} = \frac{g_{0(1)} - \cos k}{\epsilon_k^{0(1)}}
\end{eqnarray}
Following the procedure outlined in Eq.~(\ref{eq:ukn}) and the subsequent paragraph, we obtain:
\begin{eqnarray}
u_k(n) &=& -\beta_k^* F(n-1, \xi_k)\nonumber \\
&=& \Big( -i \left( n_{kx}^0 \sin \phi_k^0 \cos \phi_k^1 + n_{kx}^1 \sin \phi_k^1 \cos \phi_k^0 \right) \nonumber \\
&& + \left(n_{kx}^1  n_{kz}^0  - n_{kx}^0 n_{kz}^1 \right) \sin \phi_k^0 \sin \phi_k^1 \Big) \frac{\sin n \theta_k}{\sin \theta_k}\nonumber\\\\
v_k(n) &=& \cos n \theta_k + i \mathfrak{Im}[\alpha_k^*] \frac{\sin n \theta_k}{\sin \theta_k}\nonumber\\ 
&=& \cos n \theta_k + i \Big( n_{kz}^1 \cos \phi_k^0 \sin \phi_k^1 \nonumber\\
&&+ n_{kz}^0 \sin \phi_k^0 \cos \phi_k^1 \Big) \frac{\sin n \theta_k}{\sin \theta_k}\nonumber\\
\label{eq:ukvksq}
\end{eqnarray}
\begin{figure}[htb]
\begin{center}
\sg[~]{\includegraphics[width=0.494\columnwidth]{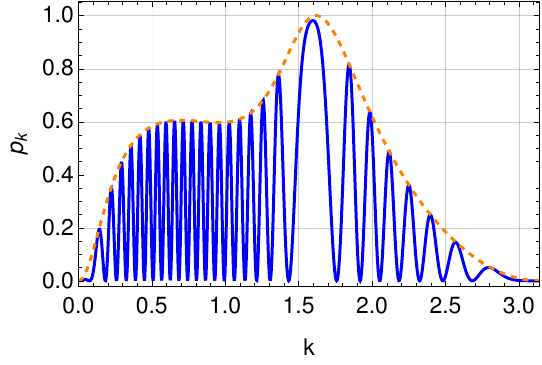}\label{fig:A1a}}
\sg[~]{\includegraphics[width=0.494\columnwidth]{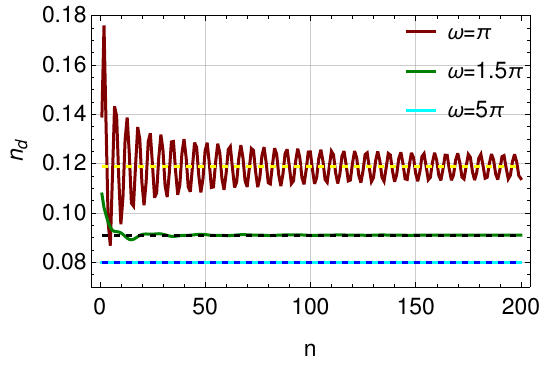}\label{fig:A1b}}
\end{center}
\caption{(a) The behavior of $|u_k(n)|^2$ as a function of $k$ for $n=30$, plotted as a solid blue line for square pulse.(b) The defect density generated after $n$ cycles is plotted against $n$ for different frequencies for square pulse. The dashed lines represent the saturation value for large $n$. Parameters used: $L=2048$, $g_0=2$, and $g_1=0.5$.}
\label{fig:A1}
\end{figure}
Starting with the initial state $|\psi_k(t=0)\rangle = \left(0,1\right)^{\text{T}}$ for each $k$ mode, we obtained the excitation probability:
\begin{eqnarray}
p_k(n)&=&|u_k(n)|^2\nonumber\\
&=&\Big( \left( n_{kx}^0 \sin \phi_k^0 \cos \phi_k^1 + n_{kx}^1 \sin \phi_k^1 \cos \phi_k^0 \right)^2 \nonumber\\
&+&\left(n_{kx}^1 n_{kz}^0-n_{kx}^0 n_{kz}^1  \right)^2 \sin^2 \phi_k^1 \sin^2\phi_k^0\Big) \frac{(1-\cos 2n\theta_k)}{2 \sin^2 \theta_k}\nonumber\\
\label{eq:pkns}
\end{eqnarray}
Using the above Eq.~(\ref{eq:pkns}), defect density for the square pulse protocol is calculated by summing over all momentum modes in the Brillouin zone using the definition Eq.~(\ref{eq:def_def}). In Fig.~\ref{fig:A1a}, we plot the excitation probability calculated using the above expressions and show similar behavior as shown for the delta kick protocol. The $n$ independent part of the Eq.(\ref{eq:pkns}) form the envelop of that oscillation. Defect density after $n$th drive cycle is plotted in Fig.~\ref{fig:A1b}. Similarly, from Eq.~(\ref{eq:ukvksq}), one can derive the expressions for the residual energy, magnetization, and log fidelity by using the definition given in Eq.~(\ref{eq:res_def}), (\ref{eq:mag_def}) and (\ref{eq:fid}) from the main text.\\
In Fig.~\ref{fig:A2} we have plotted the residual energy, magnetization, and log fidelity and the behavior is similar to the delta kicked protocol. The residual energy is plotted in Fig.~\ref{fig:14a} and with initial oscillation it saturates in large $n$. Similar behavior is also observed in the case of magnetization and log fidelity in Fig.~\ref{fig:14b} and \ref{fig:14c} respectively. Magnetization as a function of drive frequency after $n=1000$ drive cycles is plotted in Fig.~\ref{fig:14c}, and dynamic freezing is observed at certain frequencies where the magnetization remains at its initial value throughout the drive. Here, we have used $g_0 = -g_1 = 20$. The dynamical freezing phenomenon can be explained easily using Eqs.~(\ref{eq:mag_def}) and (\ref{eq:pkns}). For large $g_0$, the following approximations can be made:
\begin{eqnarray}
\epsilon_{k}^{0(1)} &\simeq& \pm g_0, \\
\phi_k^{0(1)} &\simeq& \pm\frac{Tg_0}{2}.
\label{eq:sq_fr_cond}
\end{eqnarray}
Thus, for $\frac{Tg_0}{2} = m\pi$, where $m$ is a positive integer, we obtain $|u_k(n)|^2\simeq0$. From this condition, we derive the driving frequencies where freezing phenomena occur,
\begin{eqnarray}
\omega_{fr} = \frac{g_0}{m}, \quad (m = 1, 2, 3, \ldots). 
\label{eq:freezing2}
\end{eqnarray}
and the above result matches the conditions given in \cite{df2}.\\
Finally, we compute the fermionic correlation function as defined in Eq.~(\ref{eq:cor}), and plot the absolute value of its $n$-dependent part $\delta C_0(n)$ with $n$ in Fig.~\ref{fig:13a} in. At large $n$, the correlation function, like other observables, saturates to a constant value. However, the approach towards the saturation value follows a power law, characterized by an exponent of either $-1/2$ or $-3/2$ discussed in \cite{Krishentangle,Diptiman}. As shown in Fig.~\ref{fig:13a}, we observe that $\delta C_0(n)$ follows a power law with an exponent of $-3/2$. This behavior can be understood using the saddle point approximation, similar to the case of the delta kick protocol. The saddle point is found by setting the first derivative of $\theta_k$ equal to zero. The roots of the resulting equation($k=k_0$) correspond to the saddle points and yield the phase diagram, which we reproduce in Fig.~\ref{fig:13b}. There are two distinct cases for the roots that form the phase diagram.

\textbf{Case 1:} If $k_0 = 0$ or $\pi$ are the only roots, then the saddle-point approximation finds that $\delta C_0(n)$ follows a power law with respect to $n$, characterized by an exponent of $-3/2$.

\textbf{Case 2:} The saddle point occurs at any point $k_0$ except $0$ and $\pi$ and within the interval $0 < k_0 < \pi$. In this case, the value of $k_0$ depends on $g_0$ and $\omega$. We numerically solve the following equation to check whether the root \(k_0\) exists or not.
\begin{eqnarray}
&&\Big(\left(\pi\Xi_k^2 + g_0 w \Xi_k \tan \left( \frac{\pi}{w} \right) \right)\tan\left( \frac{\pi \Xi_k}{w} \right)\nonumber\\&&+\pi\Xi_k\left( 1 - g_0 \cos k \right) \tan \left( \frac{\pi}{w} \right) \Big) = 0
\end{eqnarray}
We have taken $g_1=0$ here and $\Xi_k = g_0-\cos k$. If the root $k_0$ exists, then $\delta C_0(n)$ as a function of $n$ follows a power-law with an exponent of $-1/2$. The transition between these two dynamical phases occurs at the critical frequency $\omega_c$, where the second derivative $\frac{d^2 \theta_k}{dk^2}$ is also zero. These are the stationary points and are given by the solution to the equations provided below:
\begin{eqnarray}
g_0 w \left(\cos\left(\frac{(g_0\pm 2) \pi}{w}\right)-\cos\left(\frac{g_0 \pi}{w}\right)\right)\nonumber \\\mp2(g_0\pm1)\pi \sin\left(\frac{(g_0\pm2)\pi}{w}\right)=0
\end{eqnarray}
Please note that the $\pm$ sign corresponds to orange($k_0=\pi$) and red($k_0=0$) straight lines on the right side of the phase diagram.  The above two equations are new findings. No analytical solution for the boundary of the blue islands is possible on the left hand side of the phase diagram .\\

%
%



\end{document}